# Steric Effect in Threshold Photoionization Dissociations of Serine Conformers**

Shan Xi Tian,[†]* Jinlong Yang,[†] Hai-Bei Li,[†] Yang Pan,[‡] Taichang Zhang,[‡] and Liusi Sheng[‡]

[†] *Hefei National Laboratory for Physical Sciences at Microscale, Department of Chemical Physics, University of Science and Technology of China, Hefei, Anhui 230026, China*
[‡] *National Laboratory of Synchrotron Radiation, University of Science and Technology of China, Hefei, Anhui 230029, China*

* Corresponding author. E-mail: sxtian@ustc.edu.cn.
** Electronic supplementary information (ESI) available.

## Abstract

Steric effect in the threshold dissociative ionizations of serine conformers [$CH_2OH$-$C_\alpha H(NH_2)$-$C_\beta OOH$] is revealed by high-level *ab initio* calculations combined with our newly developed infrared laser desorption / tunable VUV photoionization mass spectrometry. We find that near the ionization thresholds the $C_\alpha$—$C_\beta$ and $C_\alpha$—C bonds are selectively broken for the respective cationic conformers, yielding the different fragments. Novel dynamic processes, proton transfer and reorientation between the predissociative fragments, are involved in the threshold photoionization dissociations.

*ARTWORK*

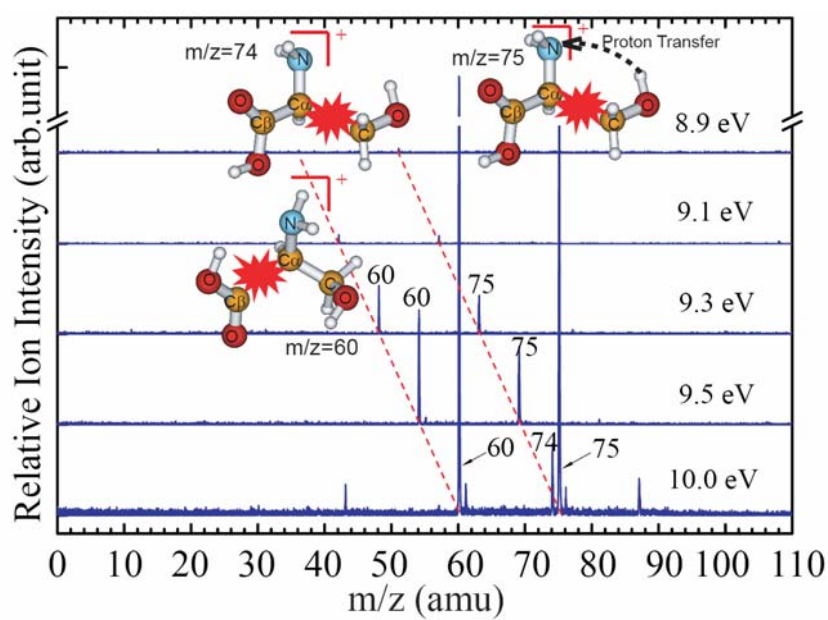

The knowledge of the formation, stabilities, and rearrangments of radical-cationic amino acids (RCAAs) is of great importance in chemistry and biology, e.g., disease development[1,2] and protein sequentiation.[3] It is a serious fact that the active RCAAs can induce more subreactions of the irradiation and oxidation damages in living cells.[2] The ultraviolet photophysics and photochemistry of amino acids can access the RCAA studies in laboratories, using various techniques, e.g., the infrared spectroscopy, microwave spectroscopy, ultraviolet photoelectron (UP) spectroscopy, and mass spectrometry.[4-11] Recently, the dissociative ionization dynamics of several amino acids was investigated both by the vacuum ultraviolet (VUV) photoionization mass spectrometry experiments[6-8] and *ab initio* molecular dynamics simulations.[12-16] However, the dissociative ionization dynamics of amino acids is too complex to be well understood due to the variety of conformations and their intramolecular hydrogen bonds (HBs).[6-8, 12-16] Here, for the first time, we report an *ab initio* study combined with mass spectrometry for serine [$CH_2OH–C_\alpha H(NH_2)–C_\beta OOH$], indicating the significant steric effect in the dissociative dynamics near the ionization thresholds of its different conformers.

Four low-lying conformers of serine (**S1**, **S2**, **S3**, and **S4**) and their cations (**S1$^+$**, **S2$^+$**, **S3$^+$**, and **S4$^+$**) are fully optimized at the B3LYP / 6-311++G(2d,2p) level and depicted in Fig. 1. Table 1 lists the relative energies ($\delta$Es) and adiabatic ionization potentials (IP$_a$s) predicted at the coupled-cluster CCSD(T) level and the vertical ionization potentials (IP$_v$s) calculated with the electron propagator theory within the third-order quasiparticle approximation.[17] All calculations are performed with Gaussian 03 program.[18] A stability order for four conformers of serine is given as **S2 > S1 > S3 > S4** according to the $\delta$E values obtained in this work. The similar orders, **S1 > S2 > S3 > S4** [9,11] and **S1 ~ S2 > S4 > S3** [10] have been predicted at the lower levels of theory, indicating a subtle dependence on the theoretical methods in predictions of the conformer stabilities with the extremely small energy differences. Whatever, the dominant existence of these four conformers in gas-phase is proved, not only by the previous studies[9-11] but also our assignment to the UP spectrum[19] with the present IP$_v$ values for valance molecular orbitals (MOs), see the supplementary Table S1 and Fig. S1 as the supporting information.

It is interesting that different carbon-carbon bonds may be broken in the cationic conformers. As shown in Fig. 1, the bond elongations with respect to the neutral are 0.005 Å

($C_\alpha$–$C_\beta$) and 0.297 Å ($C_\alpha$–C) in $S1^+$, 0.125 Å ($C_\alpha$–$C_\beta$) and 0.034 Å ($C_\alpha$–C) in $S2^+$, 0.068 Å ($C_\alpha$–$C_\beta$) and 0.079 Å ($C_\alpha$–C) in $S3^+$, and 0.006 Å ($C_\alpha$–$C_\beta$) and 0.294 Å ($C_\alpha$–C) in $S4^+$. Such significant steric effects are not reported prior to this work.

Tendency of dissociations of serine cations is revealed by the present static single-point calculations and no parent species are suspected to be survived upon ionization. It is necessary to investigate whether all cationic conformers are easily (with extremely low barrier) or spontaneously (barrier-free) dissociated, or which fragment is predominant. Both rigid and relaxed potential energy profiles in terms of carbon-carbon bond lengths are plotted in Fig. 2. The rigid profiles (the structures of the moieties except for the selected carbon-carbon bond are fixed in the energy scanning) exhibit the endothermic dissociations, while the relaxed profiles (full optimization of geometries in the energy scanning) imply that the serine cations undergo the low energy barriers less than 5 kcal/mol and their dissociations are exothermic. One can find some energy jumps shadowed in colors along the relaxed profiles. Their characteristics can be clarified by tracking the variances of cationic structures along the relaxed energy profiles. As shown in the inserted schemes of Fig. 2, the proton transfer (PT) processes occur with the elongations of $C_\alpha$–C bonds to ca. 3.0 Å, 2.5 Å, and 3.1 Å for $S1^+$, $S3^+$, and $S4^+$, respectively (also see Fig. S2 as the supporting information). These points are shadowed in cyan color at their energy profiles. The yellow area for the dissociation along $C_\alpha$–$C_\beta$ bond in $S3^+$ corresponds to a unique reorientation between $^\bullet$COOH and $(NH_2CHCH_2OH)^+$ radicals. An energy barrier ca. 3 kcal/mol should be overcome for the $C_\alpha$–$C_\beta$ bond cleavage of $S2^+$. The energetically accessible PT processes during the $C_\alpha$–C bond cleavages of $S1^+$, $S3^+$, and $S4^+$ yield the abundant $(NH_2{}^\bullet CHCOHOH)^+$ radicals ($m/z$ = 75 amu), while the $(NH_2CHCH_2OH)^+$ radical ($m/z$ = 60 amu) is produced by breaking the $C_\alpha$–$C_\beta$ bonds of $S3^+$ and $S2^+$. We think that above steric effects of carbon-carbon bond cleavages and the specific dynamic processes, i.e., PT and reorientation, during the dissociations (see the schemes inserted in Fig. 2) should be closely related to the different spatial intramolecular HBs in the serine conformers and the HB interactions of predissociative fragments. This is also in line with the well-known fact that the PT is the concomitance of the HB interaction.[4-6, 12, 15]

On the other hand, the thermodynamic calculations at the CCSD(T) level predict the most energetically favorable dissociation channels tabulated in Table 2, and the energies ($D_0$) of more possible dissociations can be found in the supplementary Table S2. Among of them, only $\mathbf{S2^+} \rightarrow (NH_2CHCH_2OH)^+$ ($m/z$ = 60 amu) + $^\bullet COOH$ ($D_0 \sim -1.35$ kcal/mol), $\mathbf{S3^+} \rightarrow (NH_2CHCH_2OH)^+ + {}^\bullet COOH$ ($D_0 \sim -5.64$ kcal/mol), or $(NH_2{}^\bullet CHCOHOH)^+ + CH_2O$ ($D_0 \sim -4.70$ kcal/mol) are exothermic, however, the relaxed energy profiles shown in Fig. 2 imply that both $C_\alpha$–$C_\beta$ and $C_\alpha$–$C$ bond cleavages seem to be exothermic. Such difference can be interpreted by the HB interactions between predissociative fragments. As shown in Fig. 2 and discussed above, these HB interactions between the fragments result in the potential energy wells where the PT and reorientation processes occur. In general, the thermodynamic data in Table 2 predict that the species with $m/z$ = 60 amu, 74 amu, and 75 amu should be predominant in the threshold ionization mass spectra.

To validate the present theoretical predictions, five threshold ionization (i.e., with the impact photon energies near the $IP_a$ values 9.02 eV, 9.35 eV, 9.50 eV, and 8.99 eV for $\mathbf{S1}$, $\mathbf{S2}$, $\mathbf{S3}$, and $\mathbf{S4}$) mass spectra of serine were recorded at the VUV photon energies 10.0 eV, 9.5 eV, 9.3 eV, 9.1 eV, and 8.9 eV with our newly developed infrared laser desorption / tunable VUV photoionization mass spectrometry (IR/VUV-PIMS).[20] In accord with the above calculations, see Fig.3, no parent cations are observed, whereas two ions ($m/z$ = 60 amu and 75 amu) are predominant in the threshold ionization mass spectra and the signal of $m/z$ = 74 amu vanishes dramatically at the lower photon energies. The former two types of cationic species are produced via the different $C_\alpha$–$C$ and $C_\alpha$–$C_\beta$ bond cleavages and a PT during breaking $C_\alpha$–$C$ bond, while the latter one corresponds to a direct $C_\alpha$–$C$ bond cleavage.

In Fig. 3, the latter three spectra with the impact photon energies 9.3 eV, 9.1 eV, and 8.9 eV are possibly of $\mathbf{S1}$, $\mathbf{S2}$, and $\mathbf{S4}$ because these photon energies are close to their $IP_a$ values: 9.02 eV ($\mathbf{S1}$), 9.35 eV ($\mathbf{S2}$), and 8.99 eV ($\mathbf{S4}$); only one feasible channel to produce $(NH_2CHCH_2OH)^+$ ($m/z$ = 60 amu) is from $\mathbf{S2^+}$, with help of the thermodynamic calculations (see Table 2). The signal intensity of this ion is slightly higher than $(NH_2{}^\bullet CHCOHOH)^+$ ($m/z$ = 75 amu), implying more abundances of $\mathbf{S2}$ in our infrared desorption beam. This is in good agreement with the present calculations of $\delta E$, but contrast to the previous results.[9,11] When the impact photon energy is lower than 9.5 eV, the signal of $m/z$ = 74 amu is almost lost, implying the efficient PT process in the dissociations on the potential energy surfaces. On the other hand, with the increase of photon energy, more dissociation channels of serine cations are open. The cationic fragments with $m/z = 87$ amu, 76 amu, and 61 amu in the 10.0 eV mass spectrum may be formed by releasing the neutral $H_2O$, $^\bullet COH$, and $CO_2$, respectively. A cationic fragment with $m/z = 43$ amu could be formed by deletion of $NH_3$ from the substantive $(NH_2CHCH_2OH)^+$ ($m/z = 60$ amu) after an intrinsic PT to $-NH_2$ group.

To the best of our knowledge, in mass spectrometry, the fragmentation to RCAAs strongly depends on the impact photon energies for aliphatic amino aicds, in particular, the smaller cationic species are predominant for the high impact photon energies.[6-8] Moreover, multistep mechanisms to produce ions[3] and excess internal energies of the parent species also account for such fragmentation. The remarkable merit of our experiments is in that nearly fragment-free mass spectra can be recorded by carefully controlling a Nd:YAG laser beam to generate intact neutral molecules and by effectively tuning photon energy to the ionization threshold at an End-Station of Hefei synchrotron radiation source.[20]

In summary, the remarkable steric effects and unique proton transfer processes in the threshold ionization dissociations of serine are predicted by the high-level ab initio calculations, satisfactorily interpreting the threshold photoionization mass spectra. The present high-level ab initio study combined with the threshold photoionization mass spectrometry can be promisingly applied to reveal the novel dissociative dynamics on the potential energy surface of the ground-state cation, and enables us to have insights into the stereochemistry of amino acids.

This work is supported by NSFC (Grant Nos. 20673105, 10775130), MOST National Basic Research Program of China (Grant No. 2006CB922000), MOST 973 Program (Grant No. 2007CB815204), and CAS (Grant No. KJCX2-YW-N07). We thank Dr. F. Qi for help in the experiments.

**Notes and references**

**Figure and Caption:**

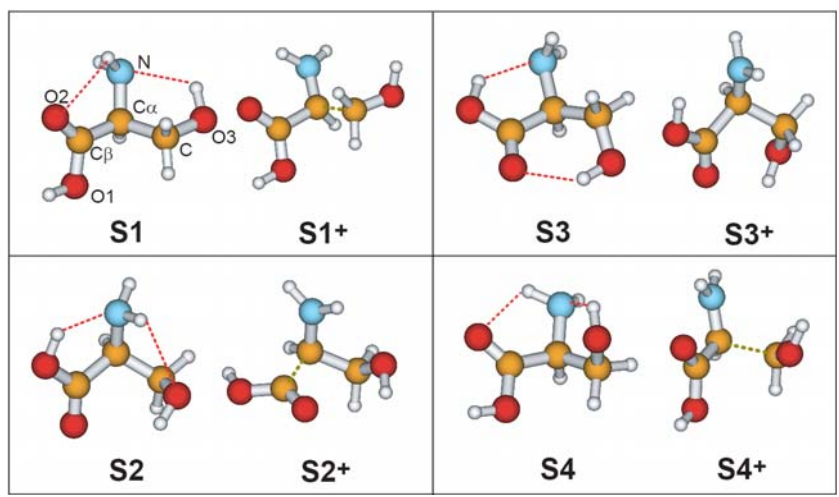

Figure 1. Serine conformers and their cations (red broken lines: intramolecular hydrogen bonds; dark yellow: potentially broken bonds)

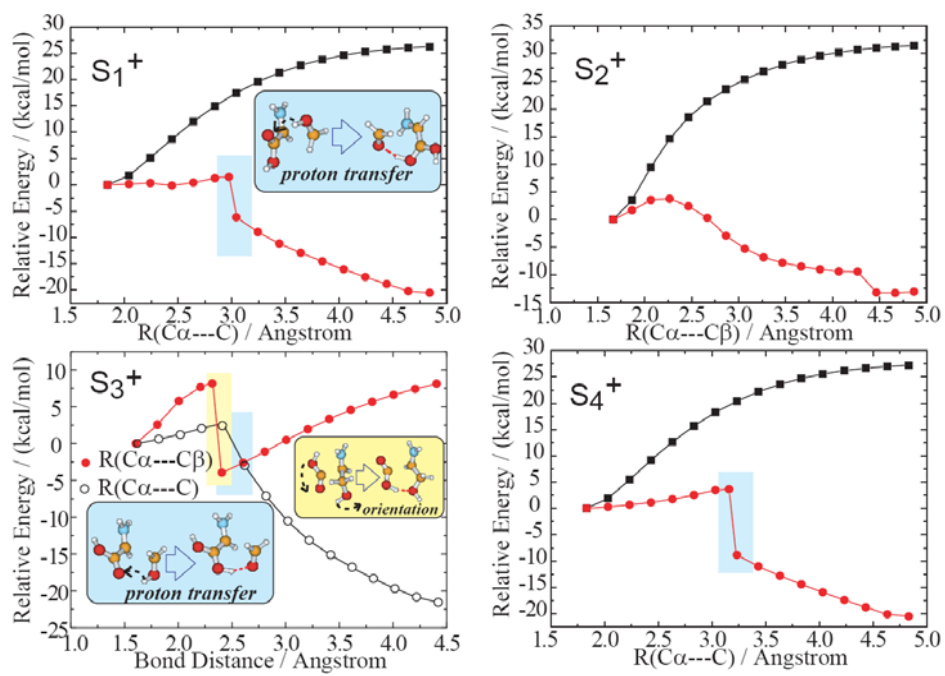

Figure 2. Energy profiles in terms of carbon-carbon distance. For **S1$^+$**, **S2$^+$**, and **S4$^+$**, ■: the rigid scanning profile, ●: the relaxed profile. Only the relaxed energy profiles are shown for **S3$^+$**.

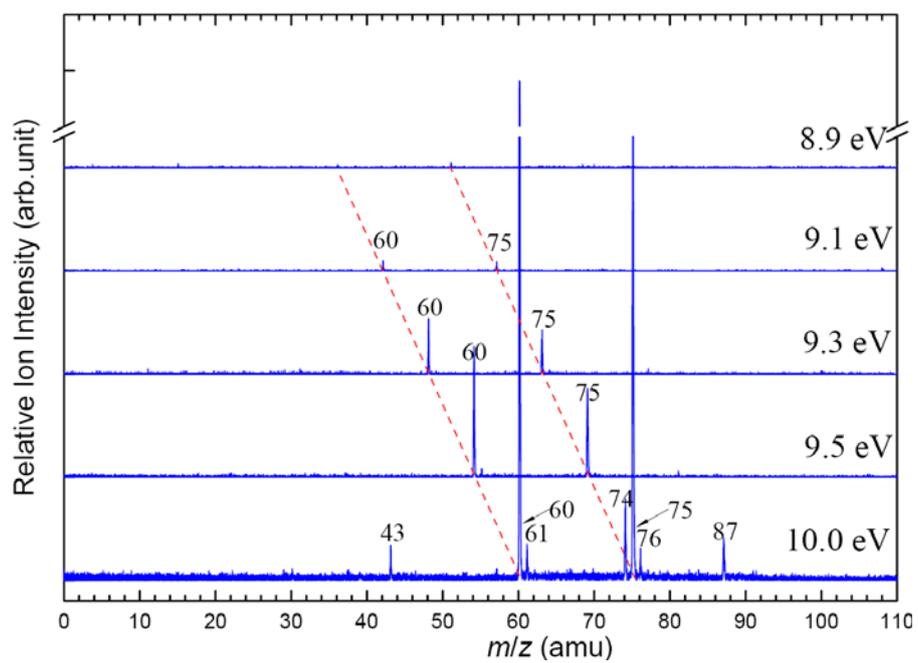

Figure 3. Threshold photoionization mass spectra of serine

**Table 1.** Relative Energies ($\delta E$ in kcal/mol), $IP_a$ and $IP_v$ Values (in eV) of Serine Conformers.

|  | **S1** | **S2** | **S3** | **S4** |
|---|---|---|---|---|
| $\delta E^a$ | 0.00 | −0.24 | 0.43 | 0.47 |
|  | (0.00) | (0.06) | (0.85) | (0.39) |
| $IP_a{}^b$ | 9.02 | 9.35 | 9.50 | 8.99 |
| $IP_v{}^c$ | 10.06 | 10.05 | 10.5 | 9.88 |

[a] The data in the parentheses were obtained at the MP2 / 6-311++ G(d,p) level.[10]

[b] The experimental $IP_a$ is estimated to be $9.1\pm0.1$ eV from the UP spectrum.[19]

[c] The experimental $IP_v$ is 10.0 eV.[19]

submittion

**Table 2. Dissociation energies ($D_0$ in kcal/mol) of serine cations.**

| Cations | Dissociations | $D_0$ |
|---|---|---|
| **S1⁺→** | $(NH_2CHCOOH)^+$ ($m/z$=74amu)+$^\bullet CH_2OH$ | 19.92 |
|  | $(NH_2{}^\bullet CHCOHOH)^{+\,a}$($m/z$=75amu)+$CH_2O$ | 3.00 |
| **S2⁺→** | $(NH_2CHCH_2OH)^+$($m/z$=60amu)+$^\bullet COOH$ | −1.35 |
| **S3⁺→** | $(NH_2CHCH_2OH)^+$($m/z$=60amu)+$^\bullet COOH$ | −5.64 |
|  | $(NH_2{}^\bullet CHCOHOH)^{+\,a}$ ($m/z$=75amu)+$CH_2O$ | −4.70 |
|  | $(NH_2CHCOOH)^+$ ($m/z$=74amu)+$^\bullet CH_2OH$ | 8.21 |
| **S4⁺→** | $(NH_2{}^\bullet CHCOHOH)^{+a}$($m/z$=75amu)+$CH_2O$ | 3.15 |
|  | $(NH_2CHCOOH)^+$ ($m/z$=74amu)+$^\bullet CH_2OH$ | 20.08 |

[a] Two conformers of this radical can be found in the supplementary Fig. S4 as the supporting information.